\documentclass[nocrop]{bioinfo}
\usepackage[english]{babel}
\usepackage{amsmath}
\usepackage{graphicx}
\usepackage[colorinlistoftodos]{todonotes}
\usepackage{upgreek}
\usepackage{float}
\usepackage{caption}
\usepackage{multirow}
\usepackage[caption=false]{subfig}
\usepackage{multicol, blindtext}

\usepackage{setspace}
\usepackage{url}

\copyrightyear{2018} \pubyear{2018}

\access{Advance Access Publication Date: Day Month Year}
\appnotes{Original Paper}

\begin{document}
\firstpage{1}

\subtitle{Gene expression}

\title[NExUS]{NExUS: Bayesian simultaneous network estimation across unequal sample sizes}
\author[Das \textit{et~al}.]{Priyam Das\,$^{\text{\sfb 1}}$, Christine Peterson\,$^{\text{\sfb 1,}*}$, Kim-Anh Do\,$^{\text{\sfb 1}}$, Rehan Akbani\,$^{\text{\sfb 2}}$  and Veerabhadran Baladandayuthapani\,$^{\text{\sfb 3,}}$}

\address{$^{\text{\sf 1}}$Department of Biostatistics, The University of Texas MD Anderson Cancer Center, Houston, TX 77030, USA, \\
$^{\text{\sf 2}}$Department of Bioinformatics \& Computational Biology, The University of Texas MD Anderson Cancer Center, Houston, TX 77030, USA, \\
$^{\text{\sf 3}}$Department of Biostatistics, University of Michigan, Ann Arbor, MI 48109, USA.
}

\corresp{$^\ast$To whom correspondence should be addressed.}

\history{Received on XXXXX; revised on XXXXX; accepted on XXXXX}

\editor{Associate Editor: XXXXXXX}

\pagebreak

\abstract{\textbf{Motivation:} Network-based analyses of high-throughput genomics data  provide a holistic, systems-level understanding of various biological mechanisms for a common population. However, when estimating multiple networks across heterogeneous sub-populations, varying sample sizes pose a challenge in the estimation and inference, as network differences may be driven by differences in power. We are particularly interested in addressing this challenge in the context of  proteomic networks for related cancers, as the number of subjects available for rare cancer  (sub-)types is often limited.\\
\textbf{Results:} 
 We develop NExUS (Network Estimation across Unequal Sample sizes), a Bayesian method that enables joint learning of multiple networks while avoiding artefactual relationship between sample size and network sparsity. We demonstrate through simulations that NExUS outperforms existing network estimation methods in this context, and apply it to learn network similarity and shared pathway activity for groups of cancers with related origins represented in The Cancer Genome Atlas (TCGA) proteomic data.\\
\textbf{Availability and implementation:} The NExUS source code is freely available for download at {{https://github.com/pdas/NExUS}}.\\
\textbf{Contact:} \href{pdas@ncsu.edu}{cbpeterson@mdanderson.org}\\
\textbf{Supplementary information:} Supplementary data are available at \textit{Bioinformatics} online.}
\maketitle


\section{Introduction}
The last decade has seen a proliferation of  large, complex datasets that quantify molecular variables such as
gene, protein, microbiome,
and population-wide genetic variation. The Cancer
Genome Atlas (TCGA) is a prime example of recent large-scale consortium-level efforts, which has generated multi-platform 'omics measurements from >10K patients across 32 common and rare cancer types. This allows for systematic investigations into the molecular mechanisms behind various oncogenic processes. Multiple studies have established that cancer initiation and progression are not outcomes of a single mutation within a gene or protein, but rather the result of a perturbation to co-ordinated networks and pathways that correspond to basic oncogenic processes such as cell proliferation and apoptosis \citep{Hanahan2000,Wang2015b}. Therefore, it is important to understand and characterize the underlying network dependency structures within and across cancers \citep{Cho2016,Hristov2017}. 
From a discovery standpoint, this is not only crucial to identify new cancer biomarkers and mechanisms, but also to distinguish the key molecular regulators of  networks in different cancers
\citep{Sonabend2014,Carro2010}.

Most existing  studies focus on identifying the characteristics of molecular networks in individual cancer populations one at a time to understand tumor-specific molecular interactions \citep{Gill2014,Creixell2015}. On the other hand, researchers are beginning to recognize the critical importance of simultaneous analysis of similar tumor types (e.g., in terms of cell origin, organ location, biological evolution) to understand fundamental commonalities and differences \citep{Weinstein2013,Tamborero2013}. 
This has led to multiple pan-cancer studies that encompass the examination of various genomic and immunologic features in related cancers such as 
different squamous carcinomas \citep{Campbell2018},  gynecologic and breast cancers \citep{Berger2018}, gastrointestinal adenocarcinomas \citep{Liu2018} and urologic cancers \citep{Chen2017}. In this article, we propose a network modeling approach for simultaneous analysis of genomic data from related tumor types.

Probabilistic graphical models \citep{Lauritzen1996} are well-established statistical tools to conduct estimation and inference of network structures. In particular, Gaussian graphical models  \citep{Whittaker1990} have gained immense popularity because of their ability to capture global dependency structures. In high-dimensional settings, such as genomics,  sparse graphical models are widely used to identify important nodes and interactions in a network consisting of a large number of genes/proteins. Various approaches for Gaussian graphical model estimation
\citep{Yuan2007, Friedman2008, Wang2012, Wang2015, Baladandayuthapani2014} have been proposed over the years.
These methods recently have been generalized  to link network estimation across multiple populations  using penalization-based \citep{Danaher2014}
or Bayesian approaches
\citep{Peterson2015}. 
The former approach is not optimal for groups with differing levels of similarity, while the latter has scaling limitations to the prior specification. Scalability in higher dimensions also is an issue for other existing methods for joint estimation of sparse graphs \citep{Guo2011, Chun2015}.  Along with learning the sparse network for each cancer type while borrowing strength across groups,  another objective of our proposed method is to avoid artefactual differences in network sparsity due to differing sample sizes.  In an application to pan-cancer network analysis, \cite{Kling2015} proposed an algorithmic method for joint network estimation with a group-specific penalty correction based on sample size. Their method, however, lacks  a formal statistical justification of the sample-size penalty term and its effect on the sparsity of the estimated networks.

 
In this article, we propose a Bayesian approach for simultaneous Network Estimation across Unequal Sample sizes (NExUS). The main advantages of the NExUS method over existing methods include (i) explicit incorporation of sample size correction in network estimation which not only allows control of the sparsity but also improves borrowing of strength between cancer-specific networks to enable a balancing of statistical power across groups; (ii) the ability to quantify 
 a {\it network similarity index} (NSI) which can be used as a global measure of network similarity among different cancer-specific networks; (iii) automatic selection of penalty parameters within a fully Bayesian framework, avoiding the need for the cross-validation step required by most  frequentist graphical modeling approaches; and (iv) an efficient strategy for updating the precision matrices, making the method more scalable both in terms of the number of variables and the number of groups than existing Bayesian methods  \citep{Peterson2015,Kundu2018}.

Our methods are motivated by and applied to The Cancer Proteomic Atlas \citep[TCPA,][]{Li2013} that has collected high-quality Reverse Phase Protein Array (RPPA) data across 32 cancer types. Many of these cancers can be divided into groups based on their histological origin, location, or similarity of biological and oncogenic processes. In this paper, we consider 4 different group of cancers, namely pan-gynecological, pan-kidney, pan-squamous, and pan-gastrointestinal. The simultaneous analysis of the proteomic networks of related cancers can help in robustly identifying the common network features. In addition, it can improve power for the discovery of features for rare cancers, which are shared with other cancers. We analyze the shared network structure of these cancers and identify shared pathway activity between all pairs of protein networks within the same group of cancers. We are able to establish both global trends in network similarity (e.g., we find that uterine carcinoma, a rare cancer, is more similar to cervical squamous cell carcinoma and endocervical adenocarcinoma, than to other gynecological cancers), as well as pathway-specific activity sharing (e.g., we find that the hormone receptor pathway has common activity across the pan-gynecological group, while the RAS/MAPK and RTK pathways have shared activation across the various types of kidney cancers).


\section{The NExUS method}

\underline{\it Data structure and notation}~~
Let $\mathbf{X}_c$ represent the $n_c \times p$ matrix of observed protein or gene expression for the $c^\text{th}$ cancer type of interest, where $c=1,2,\ldots,C$. We assume that the same set of $p$ proteins are observed for all subjects, but allow the sample sizes $n_c$ for each cancer type to differ. We assume that the data for each subject $i$ follows a multivariate normal distribution 
\begin{align*}
\mathbf{x}_{c,i} \sim N_p(\mathbf{0}, \mathbf{{\Theta}_c^{-1}}), \; i=1,\ldots, n_c,
\end{align*} 
with mean vector $\mathbf{0} \in \mathrm{R}^p$ and cancer-specific precision matrix $\mathbf{\Theta}_c$. The multivariate normal distribution is parametrized using the precision matrix $\mathbf{\Theta}_c$ (rather than the covariance matrix $\mathbf{\Sigma}_c = \mathbf{\Theta}^{-1}_c$) since there is a direct correspondence between the precision matrix and the conditional independence graph among variables. Specifically, in a Gaussian graphical model, entry $\theta_{ij}^c$ in the precision matrix $\mathbf{\Theta}_c$ is exactly 0 if and only if the corresponding proteins $i$ and $j$ are conditionally independent for that cancer type i.e.\ they are not connected by an edge within the conditional dependence network \citep{Dempster1972}.  Our statistical objectives are to learn a sparse network for each cancer type with an approach that both allows for borrowing of information across cancers, and avoids artefactual differences in network sparsity due to differing sample sizes. We now describe the prior formulation that allows us to achieve these goals.

\subsection{Hierarchical shrinkage priors for borrowing strength}
We construct a joint prior that both encourages sparsity of the precision matrices within each cancer type and similarity across the cancer types. To formulate this prior, we first divide the elements of each matrix into diagonal (D) and non-diagonal (ND) elements. Let $\theta_{ij}^c$ denote the $(i,j)^\text{th}$ element of the precision matrix $\mathbf{\Theta}_c$. Since $\mathbf{\Theta}_c$ is symmetric, the unique elements of $\{\mathbf{\Theta}_c\}_{c=1}^C$ can be partitioned into $C p$ diagonal  elements and $Cp(p-1)/2$ non-diagonal elements as follows: 
\begin{align*}
& \boldsymbol{\theta}_{D} = (\theta_{11}^1, \ldots, \theta_{pp}^1, \ldots, \theta_{11}^C, \ldots,  \theta_{pp}^C)_{Cp \times 1} \\
& \boldsymbol{\theta}_{ND} = (\theta^1_{12}, \ldots, \theta^C_{12}, \ldots,\theta^1_{(p-1)p}, \ldots,  \theta^C_{(p-1)p})_{\frac{Cp(p-1)}{2} \times 1}
\end{align*}
We put independent exponential priors with mean $\gamma^{-1}$ on the diagonal elements (i.e., on each element of $\boldsymbol{\theta}_{D}$),  and a gamma hyperprior on the parameter $\gamma$, with shape parameter $\alpha_\gamma$ and rate parameter $\beta_\gamma$.

To enable borrowing of strength while estimating sparse versions of the precision matrices for the set of cancers of interest, we introduce the  shrinkage parameters $\lambda_1$ and $\lambda_2$, where $\lambda_1$ controls the shrinkage within each cancer type and $\lambda_2$ induces similarity across the cancer types. Instead of taking the same $\lambda_1$ for each cancer type and the same $\lambda_2$  for each pair of cancers, we consider a more flexible scenario where the within-cancer and cross-cancer shrinkage parameters are dependent on cancer type. Let $\lambda_1^c$ denote the individual shrinkage parameter for cancer type $c$, $1\leq c \leq C$, and $\lambda_2^{cc^{\prime}}$ denote the cross-penalty across cancer types $c$ and $c^{\prime}$, $1\leq c < c^\prime \leq C$. Following \cite{Kyung2010}, the conditional prior of $\boldsymbol{\theta}_{ND}$ is given by \eqref{penalty_prior}. Here the first term aims to achieve sparsity within each cancer and the second term to achieve similarity across cancers.
\begin{align}
\pi(\boldsymbol{\theta}_{ND}|\sigma^2) \propto & \exp\big(-\frac{1}{\sigma}\sum_{c=1}^C\lambda_1^c\sum_{i < j}|\theta_{ij}^c|-
\nonumber\\& \qquad \qquad
\frac{1}{\sigma}\sum_{c<c^\prime}\lambda_2^{cc^\prime}\sum_{i < j}|\theta_{ij}^c-\theta_{ij}^{c^\prime}|\big)
\label{penalty_prior}
\end{align} 

\paragraph{Network similarity index:}
Taking a closer look into the role of the cross-group penalty parameter in the model, it is evident that higher values of $(\lambda_2^{cc^\prime})^2$ encourage more similarity between the networks of cancer $c$ and $c^\prime$ i.e., the higher the value of $(\lambda_2^{cc^\prime})^2$ the closer the network structure (edges) will be between two cancers (we will illustrate this in our simulations and real data analyses). Therefore within a group of related cancers, the similarity between cancer types $c$ and $c^\prime$ can be estimated by the penalty $(\lambda_2^{cc^\prime})^2$, which we term the NSI. Within each group of related cancers, we transform the NSI values of all pairs of cancers to be within the unit interval using a linear monotonic mapping such that the pair of cancers with the lowest and the highest NSI values are mapped to 0 and 1 respectively. We term these transformed  NSI values as \textit{normalized network similarity index} (NNSI).

\subsection{Incorporating sample size adjustment in the priors}
When estimating networks for multiple cancer types using existing methods, network sparsity will vary depending upon sample size, with larger sample sizes generally resulting in denser graphs (more edges). From a pan-cancer context, however, sparsity should not be dependent on the sample size -- since large cancers will overwhelm under-sampled or rare cancers. To counteract this dependency, we design priors using shrinkage parameters to mitigate the sample size effect. Since larger values of $\lambda_1^c$ encourage more shrinkage of the elements of the precision matrix, and hence more sparsity, our objective is to have a larger prior mean of $(\lambda_1^c)^2$ for the cancer types with larger sample sizes, and a smaller prior mean for rare cancer types with smaller sample sizes. Similarly, the shrinkage parameters $\lambda_2^{cc^{\prime}}$ encourage similarity between cancer types $c$ and $c^\prime$. 

To achieve these goals, we put gamma priors on the squared terms $(\lambda_1^c)^2$ with shape parameter $\alpha_1$ and rate parameter $\beta_1^c$, for $1 \leq c \leq C$, and gamma priors on $(\lambda_2^{cc^{\prime}})^2$ with shape parameter  $\alpha_2$ and rate parameter $\beta_2^{cc^\prime}$, for $1\leq c < c^\prime \leq C$. 
Following \cite{Kling2015}, we define $\beta_1^c$ and $\beta_2^{cc^\prime}$ as follows:
\begin{align*}
\beta_1^c = \frac{\beta_1}{(n_c^e)^2},\; \beta_2^{cc^\prime} = \beta_2 \bigg\{\frac{n_c^e+n_{c^\prime}^e}{2n_c^en_{c^\prime}^e}\bigg\}^2, 
\end{align*} where $\beta_1,\beta_2 > 0$.
Let $\bar{n}$ represent the average sample size across the $C$ cancer types. We define the effective sample size of the $c^\text{th}$ cancer type as $n_c^e = \bar{n}^\delta n_c^{(1-\delta)}$, where $0<\delta<1$.  Hence, we get the prior mean of $(\lambda_1^c)^2$ to be $\big(\frac{\alpha_1}{\beta_1}\big)(n_c^e)^2$, and the prior mean of $(\lambda_2^{cc^\prime})^2$ to be $\big(\frac{\alpha_2}{\beta_2}\big)\Big\{\frac{2n_c^en_{c^\prime}^e}{n_c^e+n_{c^\prime}^e}\Big\}^2$.
At $\delta = 0$,  $n_c^e = n_c$, so the prior mean depends only on the sample size for cancer type $c$, while at $\delta = 1$, $n_c^e = \bar{n}$, so the prior mean is equal across all cancer types. As $\delta$ approaches 0, relatively more sample size correction is induced by the prior.

\begin{figure*}
    \centering
    \caption{Prior means of within-cancer ($\lambda_1^2$) and cross-cancer ($\lambda_2^2$) shrinkage parameters as a function of $\delta$. The (a) within-cancer and (b) cross-cancer shrinkage parameters have been shown for a scenario with 3 cancer types A, B, C with sample sizes $n_A=50,n_B=100,n_C=200$.}
    \label{delta_value_variation}
\subfloat[Prior mean of $\lambda_1^2$]{
    \includegraphics[width=.48\textwidth]{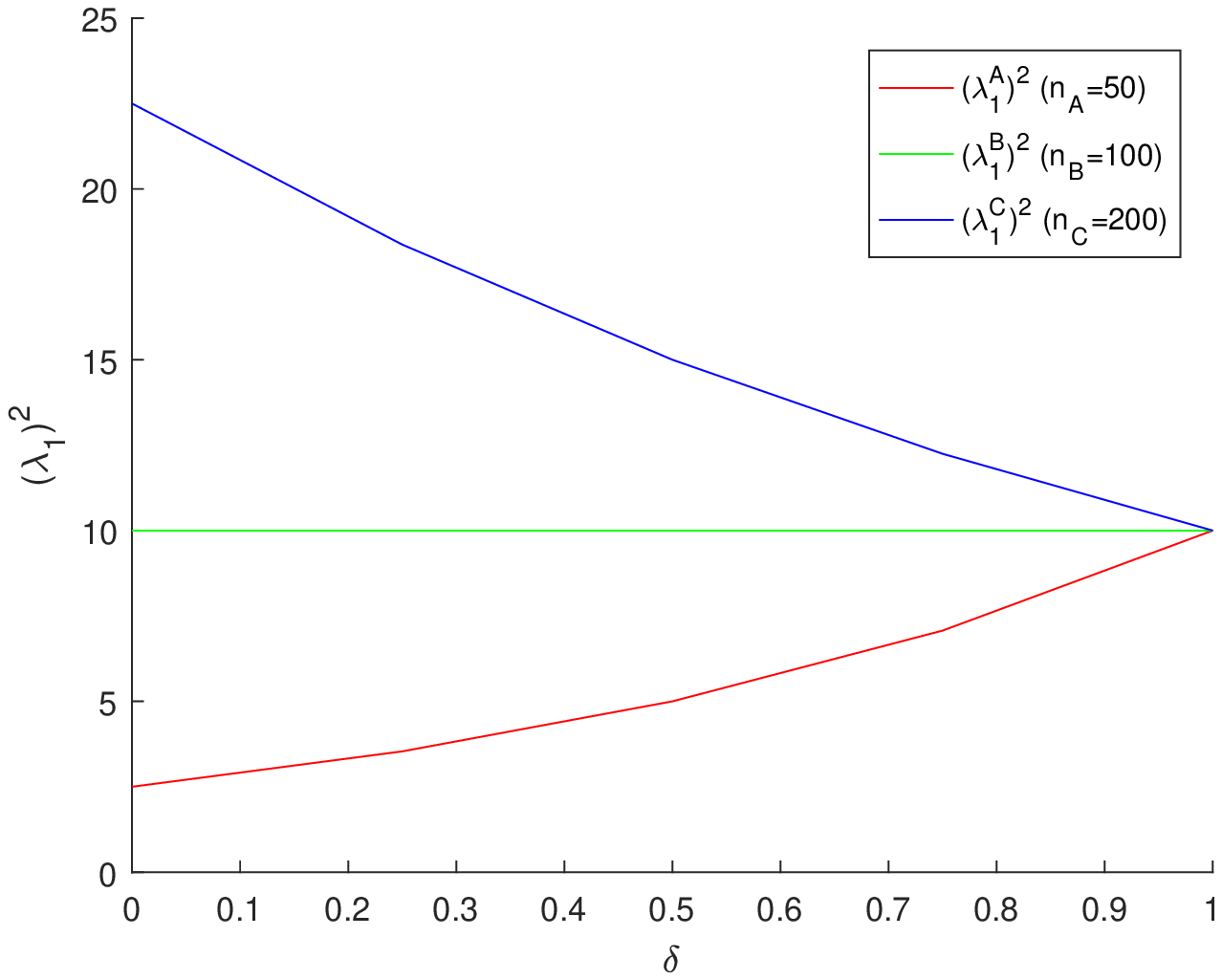}
    \label{prior_mean_1}} 
\subfloat[Prior mean of $(\lambda_2^2)$]{
    \includegraphics[width=.48\textwidth]{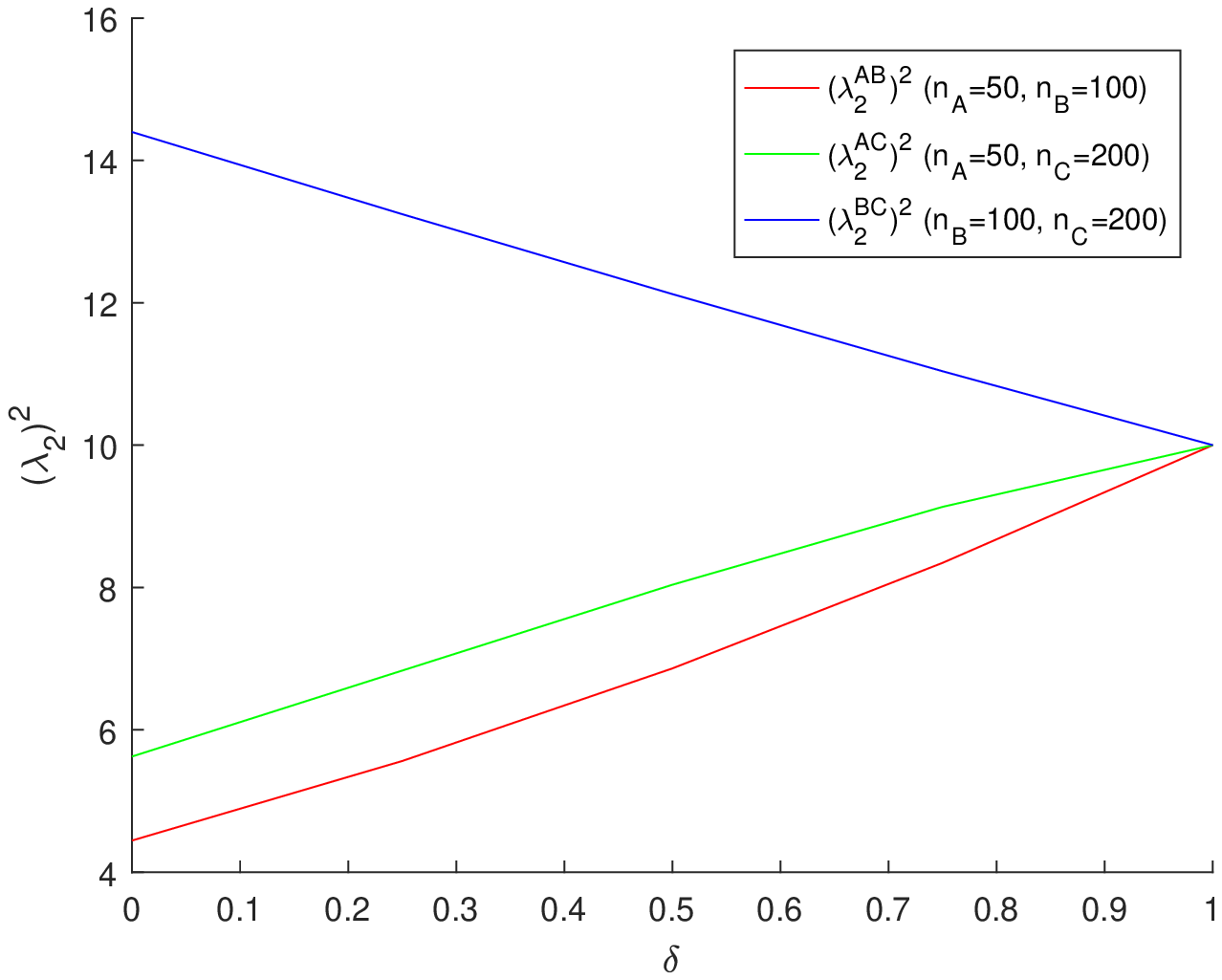}
    \label{prior_mean_2}}
\end{figure*}

\paragraph{An illustrative example:} 
Consider three cancer types A, B, and C with sample sizes $n_A = 50$, $n_B = 100,$ and $n_C = 200$. Then for $\delta<1$, our objective is to have prior mean of  $(\lambda_1^A)^2$  to be the smallest and the prior mean of $(\lambda_1^C)^2$ to be the largest, while prior mean of  $(\lambda_1^B)^2$ should be in between those values. Also, for $\delta<1$, for similarity shrinkage parameters, our objective is to have the lowest prior mean among the two cancers with the smallest sample sizes (here A and B) and the highest prior mean among the two cancers with the highest sample sizes (here B and C). 
The variation of the prior means of the shrinkage parameters for this scenario is plotted as a function of $\delta$ in Figure \ref{delta_value_variation}. Note that under the proposed way of prior construction, the desired order between the prior means of the shrinkage parameters are maintained for $\delta<1$. Also note that as $\delta$ increases to 1, the penalty terms converge in both scenarios.

\subsection{MCMC and edge selection}
After introducing latent variables $\mathbf{T}^2$ and $\mathbf{\Omega}^2$, with an appropriate choice of prior, it can be shown that the conditional prior of equation \eqref{penalty_prior} can be re-expressed  as the normal distribution 
$$\mathbf{\Theta}_{ND}|\sigma^2,\mathbf{T}^2,\mathbf{\Omega}^2 \sim N_{Cp(p-1)/2}\big(\mathbf{0},\sigma^2\Sigma_{\mathbf{\Theta}_{ND}}\big),$$
where $\Sigma_{\mathbf{\Theta}_{ND}}$ can be formulated in terms of $\mathbf{T}^2$ and $\mathbf{\Omega}^2$ (see supplementary material Section S1). This property is important as it allows us to construct a computationally efficient Gibbs sampler.

Since the posterior distribution is not tractable, we rely on Markov Chain Monte Carlo (MCMC) sampling to obtain a sample from the posterior distribution. Specifically, we are able to construct a Gibbs sampler using the posterior conditional distributions for each parameter or set of parameters. Here is a high-level outline of the updates that take place in each iteration of the algorithm. For additional details on the joint posterior, posterior full conditional distributions, and sampling steps, see Sections S2--S4 in the supplementary material. 
\begin{itemize}
\item Update the $i$th column of $\mathbf{\Theta}_c$ following the approach in \cite{Wang2012} by sampling a scalar from a gamma distribution and a vector from a multivariate normal, and then applying an appropriate transformation for $c = 1, \ldots, C$ and $i = 1, $\ldots$, p$.
\item Update the latent variable $\mathbf{T}^2$ by sampling the element $(\tau_{ij}^c)^{-2}$ from an inverse-gamma distribution for $c = 1, \ldots, C$ and $1 \leq i < j \leq p$.
\item Update the latent variable $\mathbf{\Omega}^2$ by sampling the element $(\omega_{ij}^{cc'})^{-2}$ from an inverse-gamma distribution for $1 \leq c < c' \leq C$ and $1 \leq i < j \leq p$.
\item Update $(\lambda_1^c)^2$ by sampling from a gamma distribution for $c = 1, \ldots, C$.
\item Update $(\lambda_2^{cc'})^2$ by sampling from a gamma distribution for $1 \leq c < c' \leq C$.
\item Update $\gamma$ by sampling from a gamma distribution.
\end{itemize}
Since the shrinkage prior results in sampled values for the precision matrices that have entries, which are very close to 0, but not exactly 0, thresholding is required to represent the sampled precision matrices as sparse. Therefore, to identify selected edges in the estimated networks, we choose a cut-off $\kappa = 0.05$ on the elements of the partial correlation matrices (derived from corresponding precision matrices). Instead of simply thresholding the elements of the posterior mean partial correlation matrices, we consider a slightly different probabilistic approach. From the posterior sample, we specifically calculate the posterior probability of each non-diagonal entry of the partial correlation matrices being greater than the cut-off $\kappa$. The edges corresponding to the entries with posterior probability of inclusion greater than 0.5 across all MCMC iterations are then selected (see supplementary material Section S5 for additional details on the edge selection procedure).

\begin{methods}
\section{Simulation studies}
In this section, we compare the performance of the proposed method to several existing methods on a simulated data set. 
To construct the simulated data, we consider a version of the set-up on $p = 20$ variables and $C = 4$ groups from section 5.1 of \cite{Peterson2015}, modified so that all of the graphs have at least some shared structure and there are unequal sample sizes across groups. Specifically, the first precision matrix $\mathbf{\Theta}_1$ is constructed so that all its diagonal entries are 1, $\theta_{i,i+1}=\theta_{i+1,i} = 0.5$ for $i=1,\ldots,19$, $\theta_{i,i+2}=\theta_{i+2,i} = 0.4$  for $i=1,\ldots, 18$, and all remaining entries are set to $0$. To construct $\mathbf{\Theta}_2$, 5 non-zero entries are randomly replaced by $0$ out of the 37 non-zero entries of $\mathbf{\Theta}_1$, and 5 null elements of $\mathbf{\Theta}_1$ are replaced by values sampled uniformly from $\{[-0.6,-0.4] \cup [0.4,0.6]\}$. To construct $\mathbf{\Theta}_3$, 10 edges among the shared edges of $\mathbf{\Theta}_1$ and $\mathbf{\Theta}_2$ are removed, and 10 new edges which are not present in either of those two matrices are added in the same manner. To construct $\mathbf{\Theta}_4$, 5 edges among the 22 common edges of $\mathbf{\Theta}_1$, $\mathbf{\Theta}_2$, and $\mathbf{\Theta}_3$ are removed, and 5 new edges are added that were not present in any of the first three matrices. To make sure the derived precision matrices are positive definite, we divide each off-diagonal element by
the sum of the off-diagonal elements in its row, and then average the matrix with its transpose \citep[similar to the approach taken in][]{Danaher2014}. The pairwise proportions of edges shared across the 4 precision matrices varies from 0.46 to 0.86. The sample sizes are taken to be $(n_1, n_2, n_3, n_4) = (20, 40, 60, 80)$. For each cancer type $c = 1, \ldots, 4$, we then generate $n_c$ observations by sampling from the multivariate normal distribution with mean zero and  precision matrix $\mathbf{\Omega}_c$.

For the methods comparison, we consider a variety of Bayesian and frequentist methods. In the Bayesian framework, in addition to the proposed NExUS method, we consider the Bayesian graphical lasso (BGL) and Bayesian adaptive graphical lasso  \citep[BayesAdapGL,][]{Wang2012}. The values of the tuning parameters for the BGL and BayesAdapGL are the same as those considered in \cite{Wang2012}. For edge selection, however, we use the same thresholding approach as NExUS. For NExUS, we use the following hyperparameter setting: $\alpha_1 = 1, \alpha_2 = .1, \beta_1 = .1\bar{n}^2, \beta_2 = 1\bar{n}^2, \alpha_{\gamma} = 1,$ and $\beta_{\gamma} = 1$. 
For all three Bayesian approaches, we discard the first 5,000 iterations as burn-in, and use the following 15,000 iterations as the basis for inference.

In the frequentist framework, we consider both single graph and joint estimation methods. Specifically, the single graph methods we compare to are the graphical lasso (FreqGL) \citep{Friedman2008}, adaptive graphical lasso (FreqAdapGL), and the SCAD penalized lasso \citep{Fan2009}. The joint estimation methods we consider are the group graphical lasso (GGL) and fused graphical lasso (FGL) \citep{Danaher2014}. Please see Section S6 of the supplementary material for further details regarding the application of these methods. 

To compare the performance of these methods, we primarily rely on the area under the ROC curve (AUC), as it provides a single summary of network learning accuracy. In Table \ref{AUC_per_graph}, we summarize the AUC values for the various methods for each of the four simulated graph structures. The reported value is the mean of AUC values estimated from 100 simulated datasets. The standard error values are reported inside parentheses. In these results, NExUS is consistently one of the top two performing methods. FGL also performs well for groups $C_1$ and $C_2$, which have relatively smaller sample sizes and benefit most from the enforcement of similarity across graphs, but its performance is relatively worse for $C_3$ and $C_4$. Some of the single graph methods (BGL and BayesAdapGL) become competitive when applied to the group with the largest sample size, but lack power in the classes with smaller sample sizes. In Table \ref{AUC_shared_edges}, we show the accuracy of learning shared edges, as summarized by the AUC for each pair of cancer types based on 100 simulated datasets. Again, NExUS demonstrates the best overall performance, with FGL and GGL also competitive.   FGL and GGL, while achieving good performance overall, are somewhat less flexible than NExUS, as they rely a single value for $\lambda_1$ for each class and a single value of $\lambda_2$ for each pair of classes. The procedure of finding the optimal values of those penalty parameters is computationally challenging and not scalable for networks with a large number of nodes. 

\begin{table}[!t]
\centering
\caption{Mean AUC values of graph structure learning across 100 simulated data sets, with standard deviations in parentheses. The results for the top two performing methods for each cancer type are highlighted in bold.}
\label{AUC_per_graph}
\resizebox{.75\columnwidth}{!}{%
\begin{tabular}{ccccc}
\hline
Methods & $C_1$ & $C_2$ & $C_3$ & $C_4$ \\ \hline
NExUS & \textbf{0.83(0.03)} & \textbf{0.88(0.02)} & \textbf{0.94(0.02)} & \textbf{0.94(0.02)} \\
FGL & \textbf{0.85(0.02)} & \textbf{0.88(0.02)} & 0.85(0.03) & 0.79(0.03) \\
GGL & 0.80(0.03) & 0.83(0.03) & \textbf{0.90(0.03)} & 0.91(0.02) \\
BGL & 0.84(0.05) & 0.80(0.03) & \textbf{0.90(0.03)} & \textbf{0.92(0.03)} \\
BayesAdapGL & 0.82(0.01) & 0.81(0.01) & 0.89(0.01) & \textbf{0.92(0.00)} \\
FreqGL & 0.79(0.03) & 0.77(0.04) & 0.87(0.03) & 0.88(0.04) \\
FreqAdapGL & 0.68(0.06) & 0.73(0.04) & 0.84(0.03) & 0.88(0.04) \\
SCAD & 0.54(0.04) & 0.72(0.04) & 0.84(0.03) & 0.89(0.04) \\ \hline
\end{tabular}}
\end{table}

\begin{table}[!t]
\centering
\caption{Mean AUC values of shared edge learning across 100 simulated data sets, with standard errors in parentheses. The results for the top two performing methods for each cancer type are highlighted in bold.}
\label{AUC_shared_edges}
\resizebox{\columnwidth}{!}{%
\begin{tabular}{ccccccc}
\hline
Methods & $(C_1,C_2)$ & $(C_1,C_3)$ & $(C_1,C_4)$ & $(C_2,C_3)$ & $(C_2,C_4)$ & $(C_3,C_4)$ \\ \hline
NExUS & \textbf{0.83(0.02)} & \textbf{0.91(0.02)} & \textbf{0.90(0.02)} & \textbf{0.92(0.02)} & \textbf{0.92(0.02)} & \textbf{0.94(0.01)} \\
FGL & \textbf{0.89(0.02)} & \textbf{0.94(0.02)} & 0.85(0.02) & \textbf{0.93(0.02)} & \textbf{0.93(0.02)} & 0.84(0.03) \\
GGL & 0.81(0.03) & 0.86(0.02) & \textbf{0.86(0.03)} & 0.87(0.03) & 0.88(0.03) & \textbf{0.91(0.02)} \\
BGL & 0.80(0.04) & 0.84(0.04) & 0.83(0.04) & 0.83(0.03) & 0.82(0.03) & 0.89(0.02) \\
BayesAdapGL & 0.79(0.00) & 0.80(0.00) & 0.79(0.00) & 0.82(0.00) & 0.81(0.00) & 0.89(0.00) \\
FreqGL & 0.79(0.03) & 0.81(0.03) & 0.81(0.03) & 0.77(0.05) & 0.77(0.06) & 0.86(0.03) \\
FreqAdapGL & 0.68(0.06) & 0.70(0.07) & 0.7(0.08) & 0.72(0.05) & 0.71(0.05) & 0.84(0.04) \\
SCAD & 0.54(0.04) & 0.54(0.05) & 0.54(0.07) & 0.71(0.06) & 0.70(0.06) & 0.85(0.04) \\ \hline
\end{tabular}}
\label{C4_2}
\end{table}

To gain insight into how NExUS improves on separate Bayesian inference, we plot the true positive rate (TPR), false positive rate (FPR), and Mathew's Correlation Coefficient (MCC) across a range of values for the threshold $\kappa$ on the partial correlation matrix elements in the supplementary material Figure S1. This figure demonstrates that the NExUS method outperforms the Bayesian graphical lasso (BGL) across a range of $\kappa$ values corresponding to desirable values of the TPR and FPR, primarily due to a much faster drop-off in the FPR as $\kappa$ increases. Additional details on the simulation set-up and results (including ROC curves, MCC values, and sensitivity analysis) are provided in supplementary material Section S6.

\end{methods}

\section{NExUS analyses of pan-cancer proteomic data}
We demonstrate the utility of  NExUS to conduct pan-cancer analyses of proteomics data collected across various cancer types. We focus on the four sub-groups of related cancers: pan-gynecological (pan-gynae),  pan-kidney, pan-squamous, and  {pan-gastrointestinal (pan-GI). Gynecologic cancers have similar embryonic origins, since female hormones influence their development \citep{Berger2018}. The pan-gynae group consists of breast invasive carcinoma (BRCA), cervical squamous cell carcinoma and endocervical adenocarcinoma (CESC), ovarian serous cystadenocarcinoma (OV), uterine corpus endometrial carcinoma (UCEC), and uterine carcinosarcoma (UCS). Since kidney cancers originate from the cells of the outer layer of the kidney (the renal cortex), 
we consider the following cancers in the pan-kidney group:  kidney chromophobe (KICH), kidney renal clear cell carcinoma (KIRC), and kidney renal papillary cell carcinoma (KIRP). Squamous cell carcinomas (SCCs) 
arise from the epithelia of the aerodigestive and genitourinary
tracts and share some histological characteristics that can be used for predicting the site of origin, clinical behavior, cause, prognosis, or optimal therapies \citep{Campbell2018}. In the pan-squamous group, we consider esophageal squamous-carcinoma (ESCA(sq)), head and neck squamous cell carcinoma (HNSC), and lung squamous cell carcinoma (LUSC). Adenocarcinomas of the GI tract share similar endodermal developmental origins along with exposure to common insults that promote the tumor formation \citep{Liu2018}; hence the pan-GI group consists of colon/rectum adenocarcinoma (CORE), esophageal adeno-carcinoma (ESCA(ad)), and stomach adenocarcinoma (STAD).

Our proteomics data arises from The Cancer Proteome Atlas  \citep[TCPA,][]{Li2013}, which provides protein abundance data for TCGA samples,  and consists of RPPA-based quantifications using antibodies that cover functions including proliferation, DNA damage, polarity, vesicle function, EMT, invasiveness, hormone signaling, apoptosis, metabolism, immunological and stromal function, as well as other critical cellular signaling pathways
\citep{Akbani2014}. The proteins profiled allow for a focused exploration of the various functional mechanisms underlying oncogenic processes across tumor types.

A challenge in analyzing this data is that sample sizes for different cancers vary considerably, from 48 for UCS, to 879 for BRCA (see supplementary materials Table S4 for the number of samples available for each cancer type). Instead of estimating the proteomic networks for each cancer separately, joint estimation and borrowing of strength across  the networks gives us a broader picture of the similarities and dissimilarities among the cancers belonging to the same group. Specifically, the cross-group penalty terms in NExUS help in identifying some of the weak signals (edges) for the cancer networks with smaller sample sizes when those edges are shared by other cancers of the same group, and the sample size adjustment  encourages networks with more similar levels of sparsity across cancers belonging to the same pan-cancer group.
In addition to learning  the network structures per cancer type, we also are able to estimate the global proteomic network-based similarities  and the pathway-specific similarities among the cancers belonging to the same group. 


For the analysis, we use the same values of the prior parameters as in the simulation studies. We perform 20,000 iterations, discarding the first 5,000 iterations as burn-in for each pan-cancer group. We adopt the same posterior edge selection strategy as in the simulation studies. The posterior selected networks for each cancer type are shown in Figures S8--S21 of the supplementary materials. Here we focus on high-level take-aways of these pan-cancer analyses, in particular, on the global and pathway-specific similarity measures.

\begin{figure*}[!tbp]
    \centering
    \caption{Scatter plot of normalized network similarity index (NNSI) for each pair of cancers in the (a) pan-gynae, (b)  pan-kidney, (c)  pan-squamous, and (d) pan-GI groups.}
\subfloat[pan-gynae]{
    \includegraphics[width=.4\textwidth]{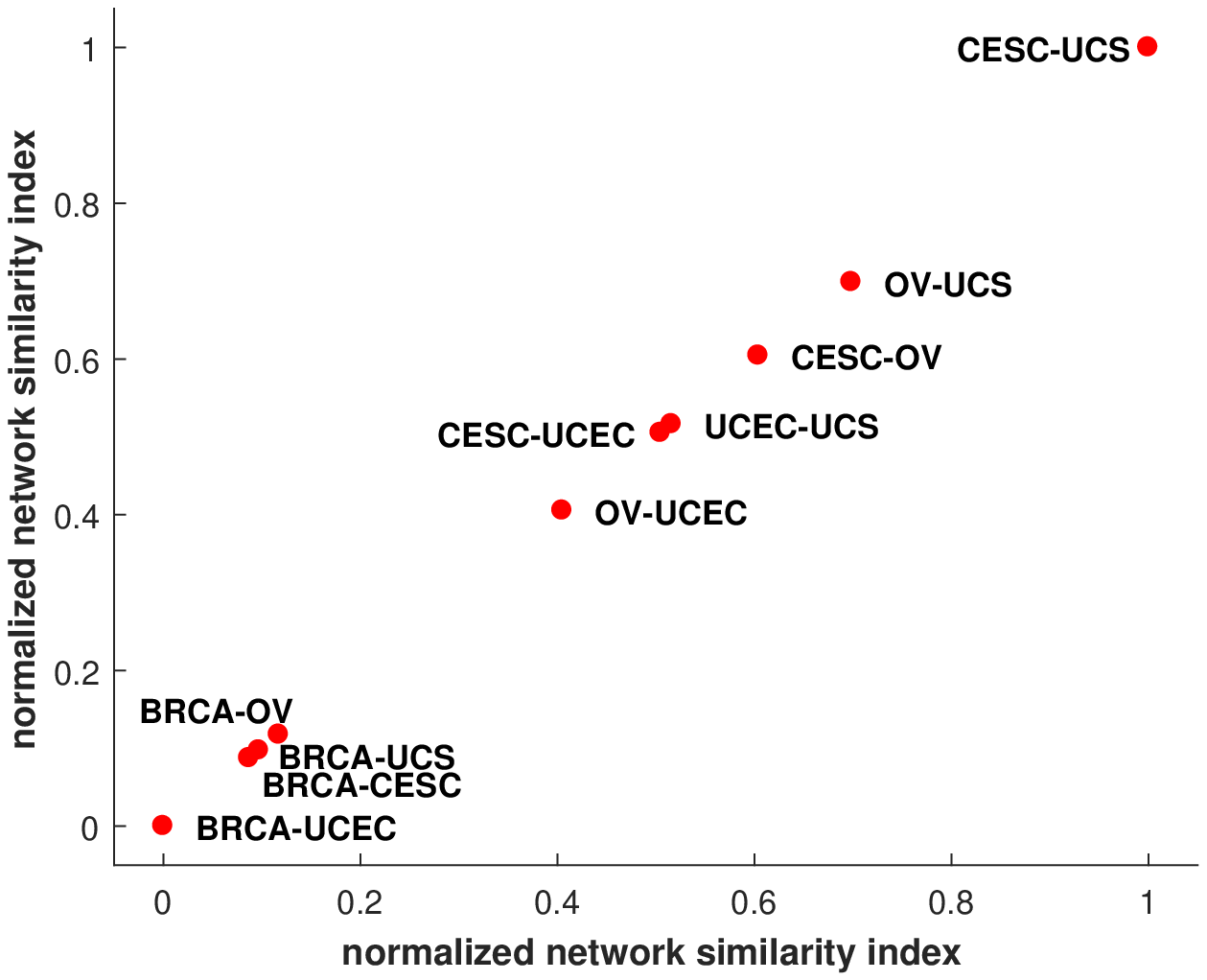}
    \label{Gyno_distance}} 
\subfloat[pan-kidney]{
    \includegraphics[width=.4\textwidth]{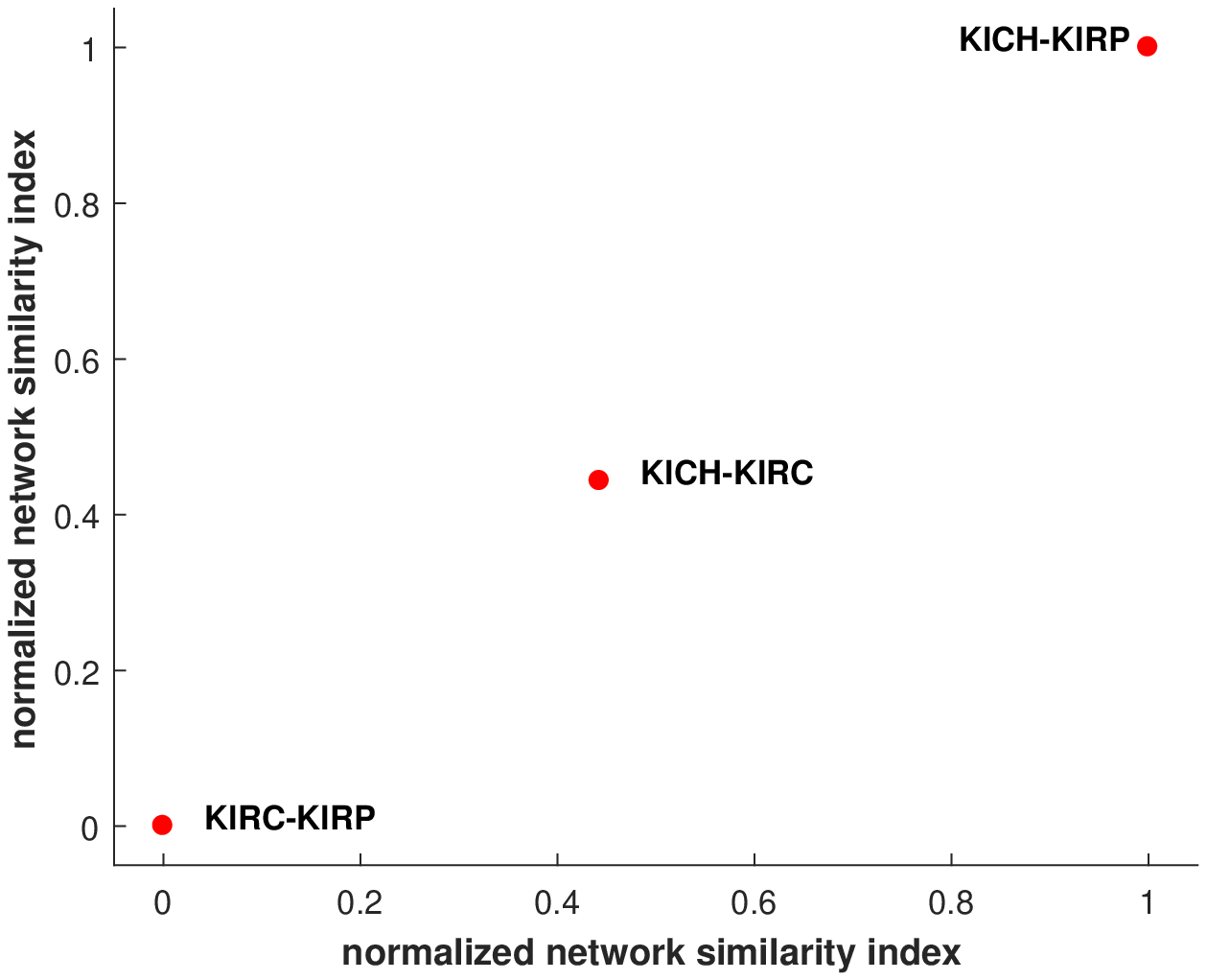}
    \label{Kidney_distance}}\\
\subfloat[pan-squamous]{
    \includegraphics[width=.4\textwidth]{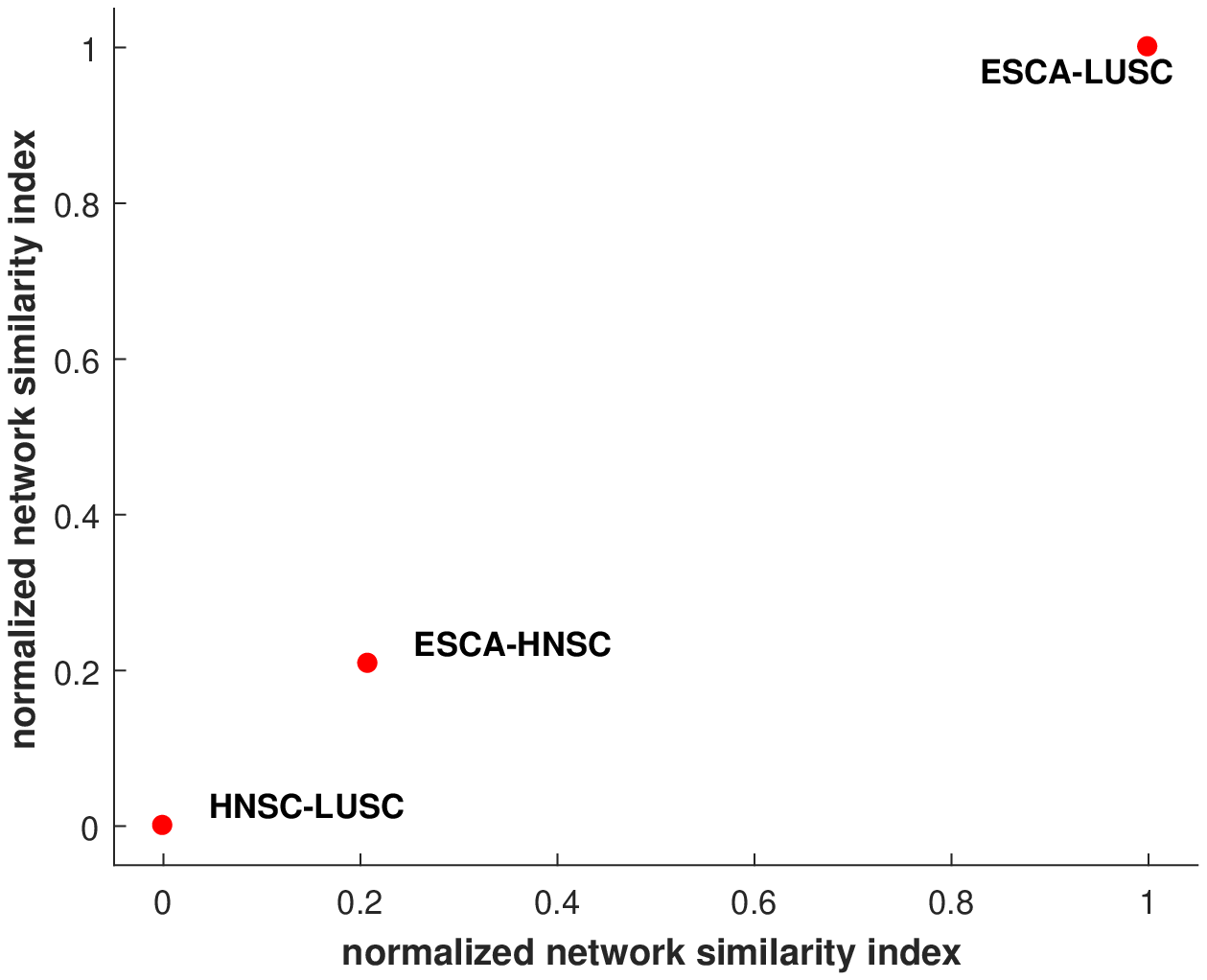}
    \label{Squamous_distance}} 
\subfloat[pan-GI]{
    \includegraphics[width=.4\textwidth]{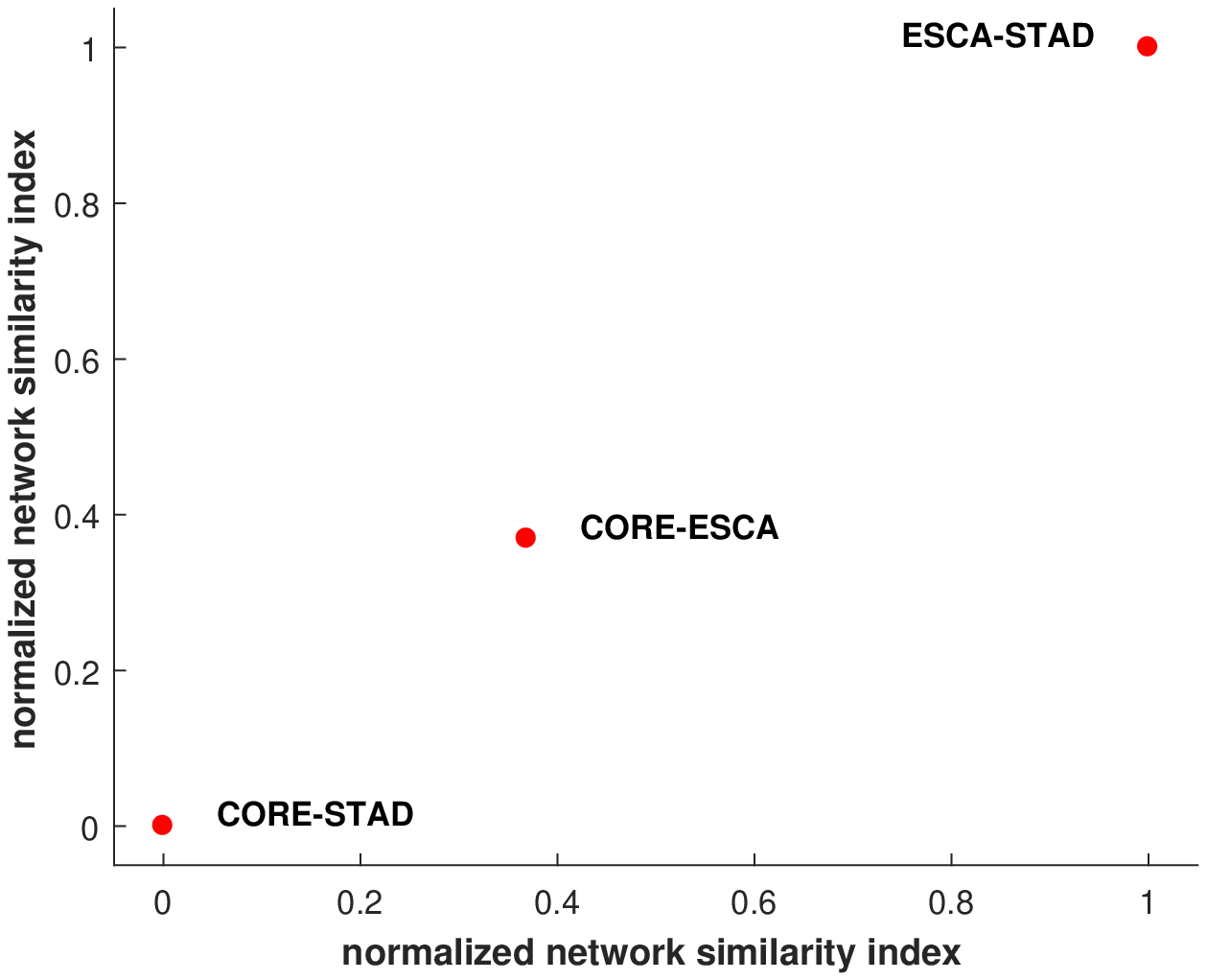}
    \label{GI_distance}}
    \label{scatter_plot}
\end{figure*}

\begin{figure}[!tbp]
    \centering
\caption{Heatmap of posterior edge probabilities for each cancer in the (a) pan-gynae, (b) pan-kidney, (c) pan-squamous, (d) pan-GI groups. For each cancer group, out of a possible $75*(75-1) / 2$ edges, we include only rows for the probabilities for edges selected in at least one graph in the group.}
    \includegraphics[width=.45\textwidth]{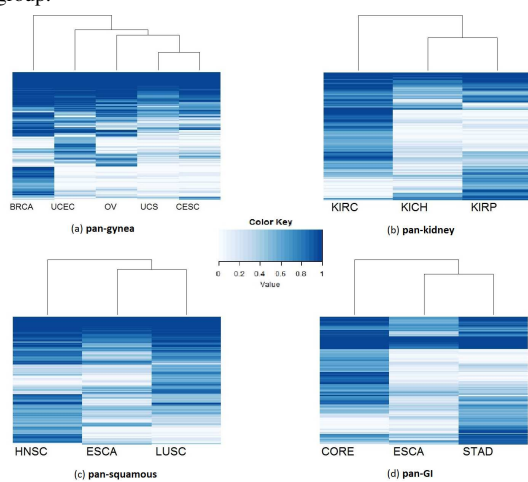}
    \label{heatmaps}
\end{figure}

\subsection{Global proteomic similarity between cancers}
To understand the extent of shared network structure among the cancers belonging to the same group, we estimate the normalized network similarity indexes (NNSI) for each pair of cancers. We plot the NNSI in Figure  \ref{scatter_plot} for all pan-cancer groups. In the pan-gynae group we observe 3 distinct clusters among the pairs of cancers based on NNSI: CESC-UCS are the closest; OV-UCS, CESC-OV, UCEC-UCS, CESC-UCEC, and OV-UCEC belong to the second cluster with intermediate NNSI values; and the third cluster, which represents the most distant pairs of cancers, contains BRCA-OV, BRCA-UCS, BRCA-CESC and BRCA-UCEC. These results show that the protein network for the rare cancer UCS is more similar to that of CESC and OV cancers as compared to other gynae cancers. BRCA has the least network similarity with other gynae cancers, suggesting BRCA has distinctive network features. In the pan-kidney group, the KICH and KIRP networks are the closest, while KIRC and KIRP networks are least similar. For the pan-squamous and pan-GI groups, ESCA(sq)-LUSC and ESCA(ad)-STAD are most similar while HNSC-LUSC and CORE-STAD are the least similar. To provide further insight into the meaning of the network similarity index, we examined the relationship between it and the $L_1$ distance between pairs of precision matrices, and found the relation to be approximately linear, where pairs of cancer with relatively high values of the network similarity index have relatively lower  $L_1$ distances between them (see supplementary material Figure S5).

Finally, to visualize the extent of shared structure across each group of cancers, we construct heatmaps of the posterior edge inclusion probabilities (Figure \ref{heatmaps}). The ordering of the cancer types, which was determined by hierarchical clustering of the edge probabilities, supports the NNSI results, with pairs of cancer scoring higher on the NNSI grouped more closely within the estimated hierarchy.

\subsection{Pathway-specific similarity between cancers}
To understand important aspects of the network similarity between cancer types, we conducted a deeper investigation of the proteomic networks using available pathway information. 
We focus our exploration on 12 well-established curated pathways with translational relevance \citep{Akbani2014,Ha2018}: apoptosis, cell cycle, DNA damage response, EMT, hormone receptor, hormone signaling breast, PI3K/AKT, RAS/MAPK, RTK, TSC/mTOR, breast reactive, and core reactive. A few of the proteins profiled are shared by more than one pathway (see supplementary material Table S5). One of our main motivations for deconvolving proteomic activity into pathways is to investigate which pathways are activated in each of the cancer types, and thereby gain a better understanding of the mechanistic and regulatory sharing of information between proteomic pathways.

To quantify pathway-specific activation, we estimate the proportion of shared edges within each pathway and across each pair of pathways for all cancers belonging to the same group. The heatmap for the proportion of shared edges within and across pathways for each pair of cancers is given in Figures S6--S7 of the supplementary materials. In general, the proportion of shared edges are higher within pathways than across different pathways, which is along expected lines. This makes biological sense, as core pathway activities are likely to be preserved across different cancer lineages. To identify the strongest shared pathway activity for each pair of cancers within each group, we computed the percentage of shared edges within each pathway for each pair of cancer types. In Figure \ref{network_distance}, we include edges based on the ranking of these percentages of edges shared: specifically, we include a colored link between cancer types in the figure for the $40\%$ highest sharing percentages across all pathways and cancer pairs for all cancer types, except pan-gynae. As the pan-gynae group has a larger number of cancer types included, we focus on the top $20\%$ there to improve legibility.

For the pan-gynae group, the hormone receptor pathway and the core reactive pathway are the most active shared pathways across the pairs of cancers. The hormone receptor pathway is  important for OV cancer, where hormonal-based systemic therapies are used to treat  ovarian stromal tumors \citep{Dacheux2013}. As UCS is a rare cancer, it is challenging to pinpoint its active pathways and to draw network-based inference. We are able to identify the cell cycle pathway as one of the top three actively shared pathways between UCS and UCEC. This is clinically relevant as a cell cycle pathway inhibitor has been identified as one of the therapeutic options for UCS \citep{Cherniack2017}. Finally, PI3K/AKT pathway activity is shared between the BRCA and UCEC as well as BRCA and OV cancer types. \cite{TCGA2012} mentions many components of PI3K pathway were amplified in basal-like breast cancers.

\begin{figure}[!tbp]
    \centering
    \caption{Shared pathway activity: Proportion of shared edges within pathways across all pairs of cancers, with edges denoting top $20\%$ (for gynae) and $40\%$ (for other cancer groups) of all pathways for each pair of cancers within each group are plotted for (a) pan-gynae, (b) pan-kidney, (c)  pan-squamous, (d) pan-GI cancer groups.}
  \includegraphics[width=.45\textwidth]{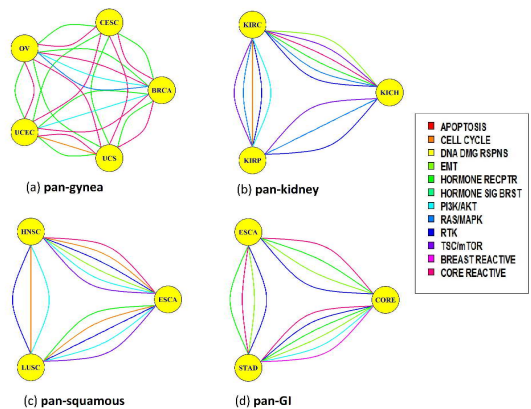}
  \label{network_distance}
\end{figure}


For the pan-kidney group, our analysis identifies the TSC/mTOR, RAS/MAPK, and RTK pathways as the top 3 actively-shared pathways. We also observe shared pathway activity for the PI3/AKT, EMT, core reactive, cell cycle, and hormone receptor pathways. \cite{Chen2016} and \cite{TCGA2013} recognized the PI3K/AKT and mTOR pathways to be active for renal cell carcinoma (RCC) and clear cell renal cell carcinoma (ccRCC). 
In addition, \cite{Gibbons2018} found EMT as one of the important activated pathways in kidney cancers.

For the pan-squamous group, the cell cycle, RTK, and PI3K/AKT pathways are activated between all pairs of cancers. Other activated pathways are TSC/mTOR, EMT, core reactive, and hormone receptor. \cite{TCGA2015} and \cite{TCGA2012a} described the PI3K, RAS, and AKT pathways as major pathways influencing HNSC and LUSC, respectively. \cite{TCGA2015} and \cite{TCGA2012a} also reported that the cell cycle and RTK pathways play an important role in HNSC and LUSC. Finally, for the pan-GI group, we find the hormone receptor, RTK, EMT, and core reactive pathways as the top 3 activated shared pathways. 

\section{Conclusion and Remarks}
In this article, we propose NExUS, a fully Bayesian method for estimating a group of related networks using Gaussian graphical modeling. The incorporation of a penalty on dissimilarity across the precision matrices in the prior specification enables borrowing of strength across the networks. This aspect of the model is especially helpful for identifying edges in the networks with smaller sample sizes. In addition, the resulting NSI helps order the relative closeness of each pair of networks within the set analyzed. Another novel feature of NExUS is the inclusion of a sample size correction which encourages similar sparsity levels across networks with different sample sizes. Our simulation studies show that NExUS outperforms other existing methods for individual and joint estimation of networks. NExUS is motivated by the TCGA-based RPPA dataset, wherein we estimate the pan-cancer proteomic networks for 4 groups of related cancers lineages. The existence of rare cancers (e.g.\ UCS in pan-gynae, and ESCA(ad) in pan-GI) makes the proposed model particularly appropriate for this application.

We developed the NExUS model under Gaussian assumptions and applied it for datasets containing a moderate number of proteins obtained using targeted profiling. In the future, it can be modified for skewed data and discrete variables, potentially following the methods proposed in \cite{Bhadra2018}. The method would then be applicable to  non-normal data sets such as mutation or copy number variation. Finally, more efficient algorithms could be developed to allow NExUS to scale to a larger number of variables, enabling an application to data sets such as the Clinical Proteomic Tumor Analysis Consortium (CPTAC), which was acquired using untargeted proteomics, and therefore has a much higher dimensionality.

\section*{Funding} 
This work was supported by the National Institutes of Health [5P30CA016672-42 (Cancer Center Support Grant, Biostatistics Shared Resource Group) to C.P., K.D., and V.B.]; [R01-CA194391, R21CA220299-01A1, R01-160736 to V.B.]; [DMS1463233 (National Science Foundation grant) to P.D and V.B.]; [CCSG  CA016672, SPORE CA140388, EDRN CA086368, CCTS TR000371, CPRIT RP160693 MD Anderson institutional Moonshot funding to K.D.]; [NIH/NCI: U24 CA210950, U24 CA210949, P30 CA016672; DoD/CDMRP: W81XWH-16-1-0237 to Akbani].

\bibliographystyle{natbib}
\bibliography{bibli}
\end{document}